# Designing AI for Online-to-Offline Safety Risks with Young Women: The Context of Social Matching

**Douglas Zytko**
Oakland University
Rochester, MI, 48309, USA
zytko@oakland.edu

**Hanan Aljasim**
Oakland University
Rochester, MI, 48309, USA
hkaljasi@oakland.edu

## Abstract

In this position paper we draw attention to safety risks against youth and young adults that originate through the combination of online and in-person interaction, and opportunities for AI to address these risks. Our context of study is social matching systems (e.g., Tinder, Bumble), which are used by young adults for online-to-offline interaction with strangers, and which are correlated with sexual violence both online and in-person. The paper presents early insights from an ongoing participatory AI design study in which young women build directly explainable models for detecting risk associated with discovered social opportunities, and articulate what AI should do once risk has been detected. We seek to advocate for participatory AI design as a way to directly incorporate youth and young adults into the design of a safer Internet. We also draw attention to challenges with the method.

## Introduction

The unique combination of computer-mediated and face-to-face communication gives rise to significant risk of harm against youth and young adults. Examples from popular media include doxing (the unwanted release of one's personal information, which can lead to online harassment and physical safety risks [17]) and swatting (pranks against Twitch live streamers and online gamers that involve calling the police with false reports in hopes of triggering a SWAT team dispatch – at least one case has resulted in a fatality [2]).

A context of online-to-offline risk to youth and young adults that we focus on in this position paper is mobile social matching apps [18], such as Tinder, Bumble, and Grindr. As of 2020, 48% of adults under the age of 30 have used a social matching app [1]. These apps enable users to discover strangers in their geographic vicinity, interact with them online, and then meet face-to-face. While once known predominantly for dating and sex, their use and design have expanded to all aspects of social interaction including friendship, activity partners, and even employment [9,13,19,20], making them a predominant avenue for young people to construct their social lives.

Research has repeatedly connected social matching app-use with online and offline harms, including sexual harassment through online messaging, and sexual assault during subsequent face-to-face meetings [1,5,7,8,14–16,22]. Safety has seldom been a motivating force behind social matching app design, however. AI serves as an valuable material for detecting and mitigating risk, and it has long served as a core component of social matching apps in the form of user-matching algorithms. Yet the use of AI for user *safety*, rather than user discovery, is rare.

Our work looks into how AI can be incorporated into social matching apps for user safety across online and in-person interaction. In this position paper we review some of our ongoing participatory AI design efforts with young women, a demographic at disproportionate risk of harm during social matching app-use. Our co-designers conceptualize roles that AI can play pursuant to risk detection in social matching apps, and build directly explainable models for how AI can detect risk in this context. We use this review to promote participatory AI design as a way to directly involve stakeholders in AI interventions for safety, and also to highlight challenges to the method.

We are committed to supporting MOSafely in any capacity that is most meaningful.

## Author Bios

Douglas Zytko is an Assistant Professor in Oakland University's Department of Computer Science and Engineering, and director of the Oakland HCI Lab. Hanan Aljasim is a PhD student researcher in the lab. As a woman of color she utilizes methods that involve direct interaction with women and other at-risk demographics to empower them in articulating their visions of safety and design trajectories that can fulfill those visions. The lab's research broadly focuses on computer-mediated risk of harm against marginalized identities. Recent work from the lab has studied computer-mediated consent to sex to understand how application design perpetuates sexual violence online and offline [23], prototypes to support women in mitigating harm in social matching contexts [24], and participatory design with woman- and LGBTQ-identifying young adults to understand their conceptualizations of safety and ideas for safety-conscious social matching apps of the future [4,6]. We typically collaborate with public health experts in the academic and non-profit sectors to inform our research.

## Participatory AI Design for Safety in Social Matching Systems

With the use and design of social matching apps expanding to ever-myriad social interaction goals, more users are likely to be exposed to harms historically perpetuated through matching app-use such as harassment, rape, and bodily harm. We are in the midst of conducting participatory design studies with young women to explore how they envision AI intervening in the matching app user experience to support their safety. Women under the age of 30 are our target demographic because they are disproportionately the victims of matching app-facilitated harm (most of our participants have been in their late teens and early 20s).

**Risk detection models:** In one such study we are supporting women in crafting directly explainable models, with supporting activities similar to those in [11], for how AI can predict the risk associated with a discovered social opportunity in a user's geographic vicinity, whether that opportunity be an individual person with a matching interest, an emergent activity, or an organized group event. See Figure 1 for an example social opportunity shown to participants to prompt their reflection on possible features in a risk detection model. The example social opportunities are tailored to the participant based on answers to a screening survey.

Participants have proposed drastically different models that reflect their varying conceptualizations of what constitutes risk. Some example factors in participants' models include crime rate at the proposed meeting location, gender of the discovered person, the number of people attending the social opportunity, and reviews from other women who have interacted with people involved in the social opportunity. Variation in participants' models suggests that social matching systems should craft user-specific models of risk detection that take into account how the user conceptualizes risk and which factors in the model they may trust more than others.

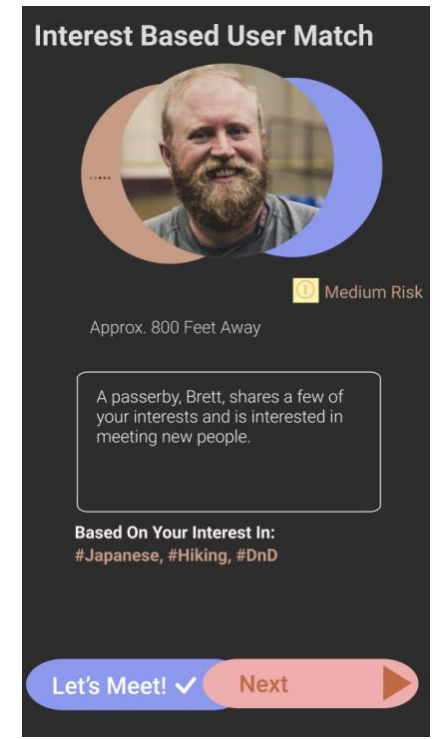

**Figure 1**: Participants are shown example social opportunities with a visible and randomly selected risk level (low, medium, or high), which is used to prompt reflection on factors they would want incorporated in a risk detection model.

**Risk has been identified…now what?:** The notion of a risk detection AI for social matching apps was borne out of a preceding participatory design study, with woman-identifying college students in their early 20s, in which participants conceptualized different roles that AI could play in the social matching app experience pursuant to safety. In addition to risk detection, two other roles pertained to what the AI should do once a social risk is detected.

One idea, dubbed the "cloaking device," refers to the AI altering a woman's visibility on the app depending on the general risk of their geographic location, and the presence of specific users in the area deemed high risk. When a woman's "cloaking device" is enabled other users would not be able to discover or interact with her on the app. This idea reflected a desire for the AI to not only detect risk, but attempt to mitigate potential harm on behalf of the user.

The other idea was dubbed a "human support network" and refers to the AI proactively alerting trusted contacts about a risky interaction (either online or offline). Trusted contacts could include friends, family members, and in some cases police. This is reminiscent to "panic button" designs discussed in prior literature [10,21], but with two key differences: the AI autonomously responds to a risky situation without the user needing to deliberately click a button, and the barrier for AI intervention is must lower. Rather than requiring that harm already have occurred, the AI would reach out to a woman's support network when the likelihood of harm occurring in an interaction reaches a threshold set by the user.

## Challenges to Participatory AI Design

Participatory design has long been heralded in its capacity to involve stakeholders in technology design [12]. Yet AI incurs unique challenges on the participatory design process, as became evident to us in our ongoing research.

One issue involves background knowledge of AI (also discussed in [3]). Most of our participants have had little to no familiarity with AI, and so we have to incorporate activities to prime them on AI possibilities such as Powerpoint presentations and interactive exercises. In addition to this taking vital time away from participants expressing their own designs, it risks biasing participants since this priming often involves us providing examples of AI and models.

Relatedly, time itself is another limitation. We divided our first participatory design study into four recurrent sessions so as not to rush participants. Yet we were still unable to proceed to model building in that time frame and needed a separate study, with mostly new participants. The capacity for participants to remain consistently involved in each stage of the design process (e.g., articulating a new AI use case > creating a model > designing an interface) is severely limited because of the time each step takes and other life responsibilities that participants have. As a result, a certain level of "translation" is needed to apply participant ideas to a subsequent design activity that they are not directly involved in.